\documentclass[runningheads]{llncs}

\usepackage[utf8]{inputenc}
\usepackage{url}
\usepackage{textcomp} 
\usepackage{tikz} 
\usepackage{underscore} 
\usepackage[]{graphicx}    
\usetikzlibrary{arrows,shadows}
\usetikzlibrary{calc} 

\begin{document}

\tikzstyle{component}=[
  draw=black, 
  thick, 
  fill=white, 
  text width=8em,
  align=center
]

\tikzstyle{multiple}=[
 double copy shadow={shadow xshift=-1ex, shadow yshift=-1ex}
 ]

\newcommand\component[5]{
  \node[component, #5](#1){
    {\large #2}\\
    {\footnotesize [#3]}\\
    \vspace{0.3cm}
    {\small #4}
  };
}

\title{Digital Availability of Product Information for Collaborative Engineering of Spacecraft}

\author{%
Diana Peters\inst{1}\orcidID{0000-0002-5855-2989} \and 
Philipp M. Fischer\inst{2}\orcidID{0000-0003-2918-5195} \and
Philipp M. Schäfer\inst{1}\orcidID{0000-0003-3931-6670} \and
Kobkaew~Opasjumruskit\inst{1}\orcidID{0000-0002-9206-6896} \and
Andreas~Gerndt\inst{2}\orcidID{0000-0002-0409-8573}}
\authorrunning{D. Peters et al.}
\institute{Software Systems for Digitalization, 
German Aerospace Center (DLR),
Mälzerstraße 3, 07745 Jena, Germany
\email{\{diana.peters,p.schaefer,kobkaew.opasjumruskit\}@dlr.de}\\ \and
Software for Space Systems and Interactive Visualization, 
German Aerospace Center (DLR),
Lilienthalplatz 7, 38108 Braunschweig, Germany
\email{\{philipp.fischer,andreas.gerndt\}@dlr.de}}
\maketitle

\begin{abstract} 

In this paper, we introduce a system to collect product information from manufacturers and make it available in tools that are used for concurrent design of spacecraft.
The planning of a spacecraft needs experts from different disciplines, like propulsion, power, and thermal. 
Since these different disciplines rely on each other there is a high need for communication between them, which is often realized by a Model-Based Systems Engineering (MBSE) process and corresponding tools.
We show by comparison that the product information provided by manufacturers often does not match the information needed by MBSE tools on a syntactic or semantic level.
The information from manufacturers is also currently not available in machine-readable formats.
Afterwards, we present a prototype of a system that makes product information from manufacturers directly available in MBSE tools, in a machine-readable way.

\keywords{Model-Based Systems Engineering \and Product Information \and Spacecraft \and Concurrent Engineering.}

\end{abstract}

\section{Introduction} \label{intro}

Complex systems can not be realized by a single person or discipline due to the systems size and heterogeneity.
The field of Systems Engineering (SE) aims to find solutions how to realize such systems successfully.
One approach emerging in the SE field is Model-Based SE (MBSE), which ``can be described as the formalized application of modeling principles, methods, languages, and tools to the entire lifecycle of large, complex, interdisciplinary, sociotechnical systems.'' \cite{Ramos2012}. 
By MBSE tool, we denote any software application that supports the development of a model by participants from different disciplines.
The spacecraft components specified in an MBSE tool we denote as Equipment, while spacecraft components built or offered by manufacturers are denoted as products.

The sources of information that engineers enter in MBSE tools for spacecraft design are PDF data sheets, spreadsheets, and engineers' implicit knowledge.
Even if information is stored and transmitted digitally, it is often not machine-readable and especially not automatically available in an MBSE tool. 
Instead, a human has to take the information from a document to enter it into another system.
This manual process is both slow and error-prone.
Furthermore, every manufacturer uses a different format and a different vocabulary to represent information.
Sometimes, not even PDF data sheets are available---Jahnke and Martelo found, that "From the 34 found suppliers of Cubesat related hardware, 62 \% do publish detailed specifications and datasheets on their website."~\cite{Jahnke2016}, which means that more than one third of the suppliers do not make this information available on their website.

As spacecraft are not anymore always one-of-a-kind products, CubeSats\footnote{CubeSats are mini satellites made of 10cm x 10cm x 10cm units and often use COTS (commercial off-the-shelf) products} and small series of satellites become more common.
According to Lange et al.~\cite{Lange2018}, it becomes also more important to reuse information from former missions.
This does not only include the MBSE models themselves, but also for example spacecraft component databases, where engineers could, additionally to the information from manufacturers, also add their own data, e.g. "template components" for different size categories of spacecraft.

Jahnke and Martelo also pointed out that "[...] the task during the CE sessions of the single domain expert is shifted from actual design of the sub-system and estimation of key-parameters towards the selection of the most suiting existing solution from e.g. a database and interface cross-check to other sub-systems."~\cite{Jahnke2016}.
So it becomes more important to find a product that fits certain requirements, including interface compatibility to other products, then to design a product from scratch.
 
In this paper we point out the problems with the current product information exchange between manufacturer and customer, collect requirements for a system to overcome these problems, and present a prototype system that makes product information available in the in-house MBSE tool of Virtual Satellite.

\section{Related Work}\label{stateart}


Several approaches exist to make the whole MBSE process or phases of it possible or easier.
This includes software applications, standards, and models of space systems.

The life cycle of a space system, as described by the European Cooperation for Space Standardization (ECSS), the European Space Agency (ESA), and the National Aeronautics and Space Administration (NASA), consists of phases 0 and A through F~\cite{ECSS2009}\cite{Nasa2008}, where in this paper we focus of the early planning phases 0 and A.
Virtual Spacecraft Design (VSD) by ESA \cite{Eisenmann2010}, Virtual Satellite by the German Aerospace Center (DLR) \cite{Fischer2017}, RangeDB by Airbus Defense and Space \cite{eisenmann2014}, and Open Concurrent Design Tool (OCDT) \cite{de2014open} as well as CDP4 \cite{CDPPAGE} by RHEA Group all aim to support an MBSE process following the suggestions made by ECSS in two Technical Memorandums, ECSS-E-TM-10-23A \cite{ECSS2011} and ECSS‐E‐TM‐10‐25A \cite{ECSS2010}. 
These tools have an internal model of the spacecraft and its subsystems. 
They all require manual input of Equipment data. 
They all support at least either the import or export of Excel spreadsheets.
Though following similar ideas and guidelines, the parameters provided by these tools, and especially the names used for those parameters, vary. Table~\ref{tab:mbseparams} shows an overview of the mass and structure parameters of most of the mentioned MBSE tools. We selected mass and structure parameters, because those are among the most relevant parameters for the early planning phases.

\begingroup
\setlength{\tabcolsep}{7pt} 
\renewcommand{\arraystretch}{1.2} 
\begin{table*}
  \centering
  \begin{tabular}{l|l|l|l}
  	\textbf{MBSE Tool}	& \textbf{Virtual Satellite 4} & \textbf{VSD}	& \textbf{OCDT/CDP4} \\
  	\hline
	\textbf{Mass} 		& massPerUnit (kg)	& Weigth [sic!] (gram) & mass (kg)\\
	\textbf{Parameters}	& 					& & mass margin \\
	\hline
	\textbf{Structure}	& radius (m)		& & diameter (m)\\
	\textbf{Parameters}	& shape				& & shape\\
						& sizeX (m)			& Height (millimetre) & height (m) \\
						& sizeY (m)			& Length (millimetre) & length (m) \\
						& sizeZ (m)			& Width (millimetre) & width (m)
  \end{tabular}
  \vspace{5pt}
  \caption{Mass and structure parameters of equipment at different MBSE tools (units in braces).
  The parameter names for Virtual Satellite 4 and VSD were taken from installations of the tools. OCDT and CDP4 directly implement the ParameterType concept from ECSS-E-TM-10-25. At Virtual Satellite, the margin parameters are not defined at Equipment level, but above.}
  \label{tab:mbseparams}
\end{table*}
\endgroup

The aim of standards and formats is to provide a base for interfaces, so systems can exchange information directly, reliably, and without human interaction. 
For the small field of Spacecraft Onboard Interface Services the Consultative Committee for Space Data Systems (CCSDS) developed a standard for Electronic Data Sheets (EDS) \cite{SANA}. 
EDS allow the exchange of information in a machine-readable format; no manual data transformation is necessary. Units, like "gram" or "inch" are defined in the EDS dictionary of terms, as are quantity kinds, like "massQK" or "lengthQK". But there is no definition that connects "gram" with "massQK" or that states that every physical component must have a mass.

ISO 10303 (STEP) \cite{Pratt2001} is an ISO standard for product manufacturing information, i.e., how a product is supposed to be produced, but it can be used for the whole life cycle of a product. 
Regarding space engineering, the technical report ECSS-E-TM-10-20A \cite{ECSS2010a} lists what part of the STEP standard should be used for information exchange between which disciplines.
This regards mostly computer-aided design (CAD) and physical structure contexts while other parts of the standard, like electronics, are neglected.
A rather simple format to store 3D information is STL (STereoLithographie, also Standard Triangulation/Tesselation Language) \cite{Roscoe1989}, which is supported by most CAD tools and used for additive manufacturing.

The specification of information exchange between manufacturers and MBSE tools does not only include technical protocols but also the information on which data is relevant in which context and what is its semantic meaning.
Tailored modeling languages and tools are required to describe the semantic model of a spacecraft.
Hennig et al. looked into existing languages and tools that can be used to describe Conceptual Data Models (CDMs).
They conclude that none of them are ideal for this task \cite{Hennig2015}. In the same year, Ait-Ameur et al. introduced a special ontology modeling language (PLIB) for engineering in general \cite{Ait-Ameur2015}.
Following their former analysis, Hennig et al. developed a conceptual data modeling language, SCDML \cite{Hennig2016b}, and also an ontology to describe space system design data \cite{Hennig2016}.
Hoppe et al. also mentioned the benefits of ontologies and Web Ontology Language (OWL) together with the Eclipse Modeling Framework (EMF) for the MBSE process \cite{Hoppe2017}\cite{Hoppe2017a}\cite{Hoppe2017b}. On top of that, they built the Semantic Engineering Modeling Framework (SEMF) \cite{Hoppe2017c}.
MARVL CIP \cite{Bieze2018} is a platform that aims to support the information exchange between agencies and manufacturers across the whole life cycle of the spacecraft.

All these approaches target at the models in the MBSE process itself and sometimes at the question of how to use the same model across different phases or how to map between models of different phases.
They should be taken into account to generate a "product data model" in a way that is compatible to the existing tools.
But none of them can be used directly to model product data.

We looked into the previously mentioned EDS \cite{SANA} and STEP \cite{Pratt2001} as potential carriers of product data. 
Both address the exchange of data sheet information, but they either focus on a small topic (EDS) or are used only for a certain area in practice (STEP---for 3D information).
So far, no standards or practices exist to cover all relevant information for the data exchange between supplier and customer in the space sector. 
However, EDS and/or STEP could become a starting point for the development of a data exchange format for product data, especially with the planned new definition of the EDS standard \cite{Prochazka2017}.

PDF data sheets are meant to describe a product technically but there is no standard regarding the syntax or semantic of this description.
There are several approaches to extract (semantic) information from data sheets, e.g. by~\cite{Castellanos1998}, \cite{Agrawal2005}, \cite{Barkschat2014}, and \cite{Murdaca2018a}, but we do not know of an accessible tool that performs that task reliably.

\section{Methods}\label{methods}

As CubeSats become more common and products for those are already offered in online shops, we compared the information offered by such shops with the information required by the above mentioned MBSE tools.
As with the MBSE tools, we focused on parameters for mass and structure.
Since the outer dimensions of CubeSats are restricted by the form factor (1U, 2U, \dots), we expected to find similar presentations of the structure parameters between the different shops.
Besides the parameter names we were also interested in the formats the information were offered in.

To use product information directly in MBSE tools, we see the following requirements for an exchange format:
\begin{itemize}
 \item \emph{machine readable} - so humans do not have to enter or copy information manually
 \item \emph{uniform / standardized} - so no transformations between different formats are necessary
 \item \emph{automatically comparable} - so search for a product that fits certain requirements becomes easier
 \item \emph{all values for one product from a single source (optional)} - so it is not necessary to request multiple sources for the information about a single product
\end{itemize}

The results of our search and the comparison with parameters at MBSE tools are discussed in the next section.

\section{Results}\label{results}

For each of the six CubeSat shops, the formats in which information was presented are summarized in Table~\ref{tab:shopformats}.
We were looking for information in tables (or bullet points in the format $<$key$>$:$<$value$>$), PDF data sheets, and STEP files.
Shop~B offered information mostly in free text\footnote{that is, running text, as opposed to bullet points, tables, or figures}, little in tables, and PDF or STEP files only upon request.
Shop~E offered parameters only in free text or in text bullet points like "Mass is ca. XYg"; Shop~D offered only very few bullet points in the format $<$key$>$:$<$value$>$ and STEP files only after login.
None of the shops offered an API (application programming interface) to read the data---we asked all of them via e-mail. One of the shops is working on an API to request the PDF data sheets, but no machine-readable data.

\begingroup
\setlength{\tabcolsep}{7pt} 
\renewcommand{\arraystretch}{1.2} 
\begin{table*}
  \centering
  \begin{tabular}{l|cccccc}
  	\textbf{Shop} 				& \textbf{A}	& \textbf{B}	& \textbf{C}	& \textbf{D}	& \textbf{E}	& \textbf{F} \\
  	\hline
	\textbf{table on website}	& X				& (X)			& X				& X				& - 			& (X)		\\
	\textbf{PDF data sheet}		& X				& -				& X				& X				& X				& X			\\
	\textbf{STEP file}			& X				& -				& -				& X				& X				& (X)		\\
  \end{tabular}
  \vspace{5pt}
  \caption{Formats of Information Presentation at CubeSat Shops}
  \label{tab:shopformats}
\end{table*}
\endgroup

For each shop, we picked the first search result for "solar panel" to compare the parameters presented directly by the shops (not in the data sheets of the manufacturers).

The mass and structure parameters of each shop are compared in Table~\ref{tab:shopparams}; Shops B, E, and F did not provide any mass or structure related parameters in a table or table-like format.

\begingroup
\renewcommand{\arraystretch}{1.2} 
\begin{table*}
  \centering
  \begin{tabular}{l|l|l|l}
  	\textbf{Shop} 		& \textbf{A}			& \textbf{C} 				& \textbf{D} \\
  	\hline
	\textbf{Mass} 		& Mass					& Very low solar cell mass	& Mass (exact mass \\
	\textbf{Parameters}	& 						& Side solar panel weight	& depends on configuration)\\
	\hline
	\textbf{Structure}	& Nominal thickness		& Solar cells thickness		& Panel Thickness\\
	\textbf{Parameters}	& Dimensions (PCB		& PCB Thickness				& \\
						&  + Solar Cells)		&							&
  \end{tabular}
  \vspace{5pt}
  \caption{Mass and structure parameters at CubeSat shops. Both parameters in Shop~D were followed by lists with the actual values.}
  \label{tab:shopparams}
\end{table*}
\endgroup

Even though our sample of CubeSat shops is small, it becomes obvious that neither between the different shops nor between the shops and MBSE tools the same parameter names are used. "Mass" and "thickness" are reoccurring names, but the additional texts at the shops (see Table~\ref{tab:shopparams}) make it clear that the semantics of those names vary (e.g., "thickness" sometimes refers to the cells only, sometimes to cells plus PCB).

\section{Prototype: Product Information in MBSE Tool}\label{results}

We built a small prototype to make product information available in MBSE tools, including a plugin for the DLR in-house MBSE tool Virtual Satellite.
Figure~\ref{fig:AO} shows an overview of the architecture.

\begin{figure*}[!htb]
\centering
\begin{tikzpicture} [auto, line/.style ={draw, thick, -latex'}]
  \matrix [row sep=15mm, column sep=16mm] {
    \component{supplier}{Manufacturers}{Software System}{Contains product data; accessible via an API}{multiple} &
    & \component{user}{Engineers}{Person}{}{multiple} \\
    \component{crawler}{Crawler}{Container}{Requests data from supplier API and creates or updates database entries}{} &
    \component{storage}{Database}{Container}{Storage of product data according to our data model}{} &
    \component{mbsetool}{MBSE Tool}{Software System}{Tool to collaboratively create the model of a spacecraft}{} \\
  }; 
  \begin{scope}[every path/.style=line]
    \draw[->] (crawler) to[bend left=20] node[midway,left] {request} (supplier);
    \draw[->] (supplier) to[bend left=20] node[midway, right] {products} (crawler);
    \draw[->] (crawler) to[bend right=20] node[midway,above] {update or}
                                                                    node[midway, below] {add entry} (storage);
    \draw[->] (mbsetool) to[bend right=20] node[midway,above] {search} (storage);
    \draw[->] (storage) to[bend right=20] node[midway, above] {matching} node[midway, below] {entries} (mbsetool);
    \draw[->] (user) to node[midway] {uses} (mbsetool);
    \draw[thick,dotted] ($(crawler.north west)+(-0.5,0.2)$) rectangle ($(storage.south east)+(0.3,-0.7)$) node[below] {\begin{tabular}{c}
                                                                                                                                                                                                      Product Data Hub \\
                                                                                                                                                                                                     {\footnotesize [ Software System]}\\
                                                                                                                                                                                                    \end{tabular}};
  \end{scope}
  
  \node[anchor=north west, yshift=-5mm] at (current bounding box.south west)
        {\begin{tabular}{ll}
            Software System:& one standalone software application\\
            Container:& data storage with an interface to store, update, delete, \\ & and search data\\
            Person:& a human being\\
          \end{tabular}
        };
\end{tikzpicture}
\caption{Architecture of Product Data System}
\label{fig:AO}
\end{figure*}

The main component is the Product Data Hub (PDH) that consists of a crawler and a database.
The database stores product information independently of manufacturers; this bridges temporary unavailability of single shops and provides a single access point for the MBSE tools.
The database is filled by a crawler that request the manufacturers regularly and updates or adds entries.
Here, we need only one crawler because all manufacturers provide their data in the same format.
To request data from actual manufacturers (given they provide an API) would require different crawlers since currently there is no uniform data exchange format all manufacturers share.

The second component of the prototype is a manually created mockup of manufacturers that provides product information in a machine-readable way via http in a JSON (JavaScript Object Notation, a language-independent data format)) format.
The data format is the same for all manufacturers.

The last component of the prototype is a plugin for Virtual Satellite that enables the users to do three different things:
\begin{enumerate}
  \item Add an \emph{Equipment} with the values of a product in the PDH
  \item Update an \emph{Equipment} with values from a product in the PDH
  \item Save an \emph{Equipment} as product to the PDH
\end{enumerate}

\begin{figure}[!htb]
  \centering
  \includegraphics[width=0.85\textwidth]{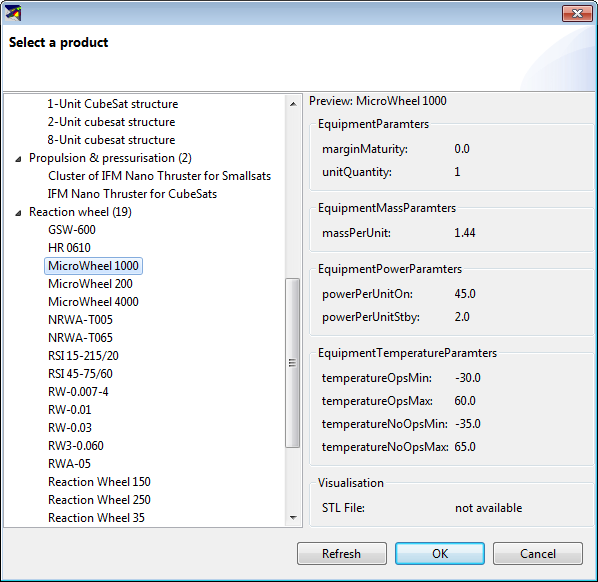}
  \caption{Screenshot of Product Selection Dialog in Virtual Satellite}
  \label{fig:dialog}
\end{figure}

To add a new \emph{Equipment} with the values of a product in the PDH, the user browses a product list provided by the plugin and selects one with fitting values (see Figure~\ref{fig:dialog}). The new \emph{Equipment} can then be used as any \emph{Equipment} in Virtual Satellite---changes at the \emph{Equipment} have no effect on the product from which the \emph{Equipment} was created.

The update function is rather a search function: 
The user specifies an \emph{Equipment} and the plugin looks for products in the PDH that fit this specification within a range of uncertainty for all values. 
The range of uncertainty can be defined by the user.
The user selects one product that fits the pre-specified \emph{Equipment} and all values from the product are taken over to the \emph{Equipment}.

The last function is to add an \emph{Equipment} as product to the PDH.
That way the user can for example define templates (e.g. for a "small battery") for which no real product from a manufacturer exists.
This function can also be used to add values for products described by PDF data sheets manually.

\section{Conclusion and Outlook}\label{conclusion}

The comparison of parameters at MBSE tools and CubeSat shops shows that there is neither a set of parameters that is supported by everyone nor a common understanding of the semantics of the provided parameters.

Our prototype shows how product information can be exchanged between manufacturers and MBSE tools in principle. 
It also shows that for machine readable exchange of product information between manufacturers and MBSE tools a standardized format is needed that includes also a semantic description for each parameter.
Our prototype only worked because we had control over all parts---to expand the concept, every party needs a common understanding of the exchanged information.
We think that over the next years it should be possible to reach such a common understanding within the space industry, at least for parts of the information to exchange.
The attempt at ESA to find a new and broader standard for EDS goes in that direction.

In the future we want to look more into semantic descriptions of space products and also in different phases of the spacecraft life cycle, since so far our focus was on the early planning phase.

\bibliographystyle{splncs04}
\bibliography{bibliography/bibliography}

\end{document}